\PassOptionsToPackage{a4paper, margin=1in, bottom=1in}{geometry}
\documentclass[pdflatex,sn-mathphys-num]{sn-jnl}% Math and Physical Sciences Numbered Reference Style 
\usepackage{graphicx}%
\usepackage{multirow}%
\usepackage{amsmath,amssymb,amsfonts}%
\usepackage{amsthm}%
\usepackage{mathrsfs}%
\usepackage[title]{appendix}%
\usepackage{xcolor}%
\usepackage{textcomp}%
\usepackage{manyfoot}%
\usepackage{booktabs}%
\usepackage{algorithm}%
\usepackage{algorithmicx}%
\usepackage{algpseudocode}%
\usepackage{listings}%
\usepackage{float}  % Add this to the preamble for using H placement
\usepackage{caption}  % Add this to customize caption width
\theoremstyle{thmstyleone}%
\theoremstyle{thmstyletwo}%

\theoremstyle{thmstylethree}%

\begin{document}

\title[Article Title]{Age-Normalized Testosterone Peaks at Series B for Male Startup Founders}


\author{\fnm{Jordan} \sur{Moradian}}
\equalcont{These authors contributed equally to this work.}

\author*{\fnm{Michael} \sur{Dubrovsky}}\email{mike@siphox.com}
\equalcont{These authors contributed equally to this work.}

\author{\fnm{Megha} \sur{Sama}}
\author{\fnm{Pavel} \sur{Korecky}}
\author{\fnm{Sidarth} \sur{Kulkarni}}
\author{\fnm{Yaniv} \sur{Goder}}
\author{\fnm{Diedrik} \sur{Vermeulen}}
\affil{\orgdiv{}, \orgname{SiPhox Health}, \orgaddress{\street{111 Terrace Hall Ave}, \city{Burlington}, \postcode{01803}, \state{MA}, \country{USA}}}


\abstract{In a study of 107 male Y Combinator founders, a surprising correlation between age-normalized testosterone and company stage was uncovered. Testosterone, a hormone associated with confidence, dominance, and drive, increased by 55.7\% from pre-seed to seed funding, peaking at the Series B stage, where levels were 99.6\% higher than pre-seed. After series B funding, testosterone was observed to drop by 42.2\%, coinciding with a spike in cortisol levels. This age-normalized biomarker analysis supports the dual-hormone hypothesis, illustrating that early startup success fosters feelings of dominance and confidence, while later-stage pressures and stresses erode these feelings. An alternative interpretation of the data, which suggests the opportunity for a longitudinal study, is that male founders with higher testosterone are more likely to raise larger rounds of funding.}




\maketitle

\section{Introduction}\label{sec1}

Testosterone is a fundamental hormone in male physiology, profoundly influencing mood, motivation, muscle mass, energy levels, and sexual function~\cite{bib1,bib4}. Its levels are dynamic and can be modulated by various psychological and environmental factors. Notably, fluctuations in testosterone have been associated with perceptions of dominance, competitive success, and social status~\cite{bib3}. Lifestyle elements such as stress, sleep quality, and physical activity also significantly impact testosterone production~\cite{bib2,bib12}.

Startup founders represent a unique population often subjected to intense work environments characterized by long hours, high stress, and irregular sleep patterns, especially during the early stages of company development. These demanding conditions could adversely affect their hormonal health. Conversely, as startups advance through funding stages —such as pre-seed, seed, Series A, and beyond—founders may experience a change in their lifestyle and self-perception, potentially influencing their testosterone levels.

We conducted a study involving 107 male and 32 female startup founders attending a Y-Combinator reunion event~\cite{bib_yc}. Participants were offered a complimentary SiPhox Health at-home test kit capable of measuring biomarkers related to hormonal health, metabolic fitness, nutritional status, cardiovascular risk, and inflammation (see Table~\ref{biomarkers})~\cite{bib_biomarkers}. The test kit utilized micro-sample blood testing technology, enabling self-collection of small finger-prick blood samples (200uL-400uL, 4-8 drops), which were subsequently analyzed in a CLIA/CAP laboratory~\cite{bib10}.

By analyzing age-normalized testosterone levels and correlating them with the funding and lifecycle stages of the founders' companies, we aimed to uncover patterns that might reflect the interplay between entrepreneurial activities and hormonal health.

\begin{table}
    \centering
    \captionsetup{width=1\linewidth} 
    \caption{The core biomarkers included in the SiPhox Health at-home test kit.}
    \label{biomarkers}
    \begin{tabular}{lcc}
        \hline
        \hline
        ApoB& Estradiol (female)& TSH\\
        ApoA1& FSH (female)& Testosterone (male)\\
        ApoB:ApoA1 ratio& Ferritin& Testosterone:Cortisol ratio (male)\\
        Total Cholesterol& HDL-C& Triglycerides\\
        HbA1c& LDL-C& Vitamin D\\
        Cortisol& Total Cholesterol: HDL-C ratio& Homocysteine\\ 
        DHEA-S& Triglycerides:HDL-C ratio& Fasting Insulin\\ 
        \hline
    \end{tabular}
\end{table}

\subsection*{Key Takeaways}

\begin{itemize} 
    \item \textbf{Correlation with Funding Stages}: Age-normalized testosterone levels in male startup founders showed a significant correlation with company funding stages. 
    \item \textbf{Peak at Series B}: Testosterone levels increased from the pre-seed stage through Series B, peaking at the Series B funding stage. 
    \item \textbf{Physiological Insights}: The hormonal patterns suggest that progression through funding stages may be associated with physiological changes in founders, potentially linked to stress reduction or enhanced feelings of success and confidence. 
    \item \textbf{Implications for Well-being}: These findings underscore hormonal health's importance in entrepreneurial success and highlight the need for strategies to support founder well-being. 
\end{itemize}

\section{Methods}\label{sec11}

\subsection{Collecting Blood Biomarker Data}\label{subsec_bbm_data}
Samples were self-collected by users via finger-prick sample collection onto an ADX100 serum separator card~\cite{bib6}. These cards were then set to a CLIA/CAP laboratory\cite{bib10} for re-hydration and processing~\cite{bib10}. All the tests offered had been previously validated as LDTs by said laboratory to have equivalence to venous draws on Beckman Immunoassay and Chemistry analyzers~\cite{bib5, bib9}. 

Given that testosterone and cortisol follow diurnal patterns, it is important to note that samples were collected over the course of eight hours in the afternoon. Test participants were blindly distributed throughout the testing period.

\subsection{Matching Company Stage Data}\label{subsec_company_stage_data}
Company stage at the time of testing was collected by matching participant email addresses to profiles on LinkedIn, PitchBook, RocketReach, or Crunchbase, finding their associated companies, then matching them to company profiles on PitchBook and Crunchbase~\cite{bib_pitchbook, bib_li, bib_cb, bib_rr}. Emails that could not be definitively matched to reliable company data were mapped to an "Unknown" company stage.\\
\subsection{Age Normalization}\label{subsec_age_norm}

To account for age-related variations in biomarker levels, we implemented an age normalization process that adjusts individual biomarker values based on age-specific reference distributions. This methodology allows for meaningful comparisons across participants of different ages by mitigating the confounding effects of age. The normalization process involves the following steps:

\begin{enumerate}
    \item \textbf{Age Binning}: Participants were divided into 5-year age bins (e.g., 20--24, 25--29, etc.) to create age groups that capture age-specific biomarker distributions.\\
    
    \item \textbf{Reference Population Sampling}: We randomly selected over 5,000 anonymized data points from the general SiPhox Health database to serve as a robust reference population for statistical analysis.\\
    
    \item \textbf{Calculation of Age Group Statistics}:
    \begin{itemize}
        \item For each biomarker, we calculated the mean and standard deviation within each 5-year age bin in the reference population.
        \item This provided age-specific statistical parameters against which individual participant data could be compared.
    \end{itemize}
    
    \item \textbf{Z-Score Transformation}:
    \begin{itemize}
        \item Each participant's biomarker value was transformed into a z-score relative to their corresponding age group's distribution in the reference population.
        \item The z-score indicates how many standard deviations an individual's biomarker value is from the mean of their age group.
    \end{itemize}
    
    \item \textbf{Global Mapping}:
    \begin{itemize}
        \item The z-scores were then mapped back to biomarker values using the overall mean and standard deviation of the entire reference population (across all ages).
        \item This step effectively standardizes biomarker values across different ages, allowing for direct comparison between participants.
    \end{itemize}
\end{enumerate}

By transforming individual biomarker measurements in this manner, we normalized for age-related differences, ensuring that observed variations in biomarker levels are more likely due to factors of interest rather than age.

\subsection{Calculating Statistical Significance}\label{subsec_statistical_significance}

To determine the statistical significance of differences in biomarker levels across company stages, we employed non-parametric statistical tests suitable for our data distribution and sample sizes. The following steps outline our methodology:

\begin{enumerate}
    \item \textbf{Data Preparation}:
    \begin{itemize}
        \item We utilized the age-normalized biomarker values obtained from the normalization process described in Section~\ref{subsec_age_norm}.
        \item Participants were grouped according to their company stage at the time of testing, as detailed in Section~\ref{subsec_company_stage_data}.
        \item Due to small sample sizes in certain stages (e.g., Series B with $n=2$ and Late Stage VC with $n=3$), we focused statistical analyses on stages with sufficient sample sizes ($n \geq 5$).
    \end{itemize}

    \item \textbf{Overall Group Comparison}:
    \begin{itemize}
        \item We performed the Kruskal-Wallis H-test to assess whether there were statistically significant differences in biomarker distributions among the different company stages~\cite{bib_kw}.
    \end{itemize}
    
    \item \textbf{Pairwise Comparisons}:
    \begin{itemize}
        \item When the Kruskal-Wallis test indicated significant differences, we conducted post-hoc pairwise comparisons using the Mann-Whitney U test (also known as the Wilcoxon rank-sum test) to identify which specific stages differed from each other~\cite{bib_mw}.
        \item The Mann-Whitney U test is a non-parametric test suitable for comparing two independent groups without assuming normal distribution.
        \item We only performed pairwise comparisons between stages with sample sizes of $n \geq 5$ to ensure the reliability of the results.
    \end{itemize}

    \item \textbf{Spearman Correlation for Trend Analysis}:
    \begin{itemize}
        \item Spearman Correlation Analysis was performed to assess the overall trend of biomarkers across ordered company stages.
        \item This non-parametric test evaluates the strength and direction of the association between the ranked variable (company stages) and the biomarker values~\cite{bib_spm}.
        \item Company stages were first converted into sorted numeric values, and then the Spearman correlation coefficient ($r_s$) and p-value were calculated for each biomarker.
    \end{itemize}
    
    \item \textbf{Handling Small Sample Sizes}:
    \begin{itemize}
        \item For company stages with very small sample sizes ($n < 5$), such as Series B and Late Stage VC, statistical tests were not performed due to insufficient power and the increased risk of Type II errors.
    \end{itemize}
    
    \item \textbf{Assumptions and Validity}:
    \begin{itemize}
        \item Non-parametric tests like the Kruskal-Wallis and Mann-Whitney U tests do not assume normality of data distributions, making them suitable for our data.
        \item We ensured that the data met the assumptions of independence and that the observations between groups were independent.
    \end{itemize}
    
    \item \textbf{Statistical Software}:
    \begin{itemize}
        \item All statistical analyses were performed using Python (version 3.11) with the \texttt{SciPy} and \texttt{statsmodels} libraries.
        \item Data manipulation and preparation were conducted using \texttt{pandas}, and visualizations were created with \texttt{matplotlib} and \texttt{seaborn}.
    \end{itemize}
\end{enumerate}

By utilizing appropriate non-parametric tests and correcting for multiple comparisons, we aimed to robustly assess the differences in biomarker levels across company stages while accounting for the limitations posed by small sample sizes and non-normal data distributions. This methodological approach allowed us to identify statistically significant differences where they existed and to report findings with an appropriate level of confidence.

\section{Results}\label{sec2}

Age distribution was controlled using anonymized aggregate data from the general SiPhox dataset to minimize confounding effects due to age-related hormonal changes. Founders in earlier funding stages were, on average, younger than those in later stages, necessitating age normalization, particularly for testosterone and DHEA-S, both of which are strongly age-dependent~\cite{bib12}. By controlling for age, we discerned trends likely attributable to stress and lifestyle factors associated with company progression rather than simply reflecting age-related hormonal decline (see Methods for details on the age-normalization process).

\begin{figure}[H]
    \centering
    \includegraphics[width=0.75\linewidth]{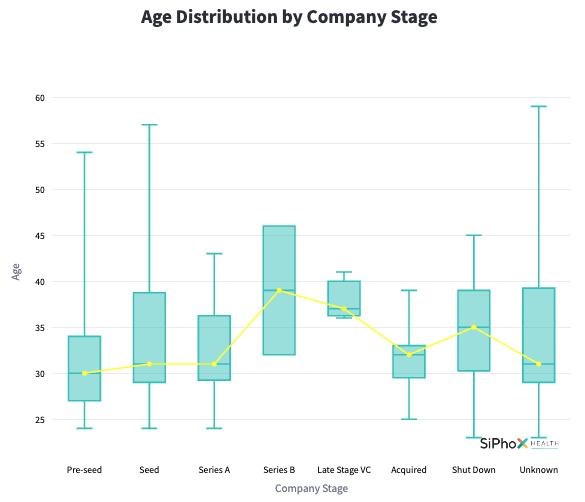}
    \captionsetup{width=0.7\linewidth} 
    \caption{Age distribution of participants across company stages. Error bars represent interquartile ranges.}
    \label{fig:cortisol}
\end{figure}

\subsection{Testosterone by Company Stage}\label{subsec2}

The age-normalized testosterone data revealed a clear trend across different funding stages (\autoref{table1}). Founders in the pre-seed stage exhibited significantly lower median testosterone levels compared to those in the seed and Series A stages. Specifically, the median testosterone level increased from 2.21~ng/mL at the pre-seed stage to 3.44~ng/mL at the seed stage, representing a 55.7\% increase, which was statistically significant ($p$\,=\,0.0478). This increase continued to the Series A stage, with a median level of 3.65~ng/mL, a 6.1\% increase from the seed stage, which was not statistically significant when compared to the seed stage ($p$\,=\,0.3635). But, the Testosterone levels peaked during the Series B stage at 4.41~ng/mL, representing a 20.8\% increase from the Series A stage; however, due to the small sample size at this stage ($n$\,=\,2), statistical significance could not be assessed.

At the late-stage venture capital (VC) phase, the median testosterone level decreased to 2.55~ng/mL, a 42.2\% decrease from the Series B peak. Similarly, the small sample size at this stage ($n$\,=\,3) was too small to calculate statistical significance. Comparing the acquired stage, with a median testosterone level of 2.56~ng/mL, we found that it was lower than the Series A stage ($p$\,=\,0.2561). The shut down stage had a median level of 3.49~ng/mL, and the difference between acquired and shut down stages was also not significant ($p$\,=\,0.4923).

\begin{table}[t]
    \centering
    \caption{Median Testosterone Levels Across Company Stages}
    \label{table1}
    \begin{tabular}{lccc}
        \hline
        \textbf{Company Stage} & \textbf{Median Testosterone (ng/mL)} & \textbf{\% Change from Previous Stage} & \textbf{Sample Size ($n$)} \\
        \hline
        Pre-seed       & 2.21 & --- & 34 \\
        Seed           & 3.44 & +55.7\% & 29 \\
        Series A       & 3.65 & +6.1\% & 11 \\
        Series B       & 4.41 & +20.8\% & 2 \\
        Late Stage VC  & 2.55 & --42.2\% & 3 \\
        Acquired       & 2.56 & +0.4\% & 6 \\ 
        Shut Down      & 3.49 & --- & 10 \\ 
        Unknown        & 3.10 & --- & 12 \\ 
        \hline
    \end{tabular}
\end{table}

\begin{figure}[H]
    \centering
    \includegraphics[width=0.75\linewidth, height=0.7\linewidth]{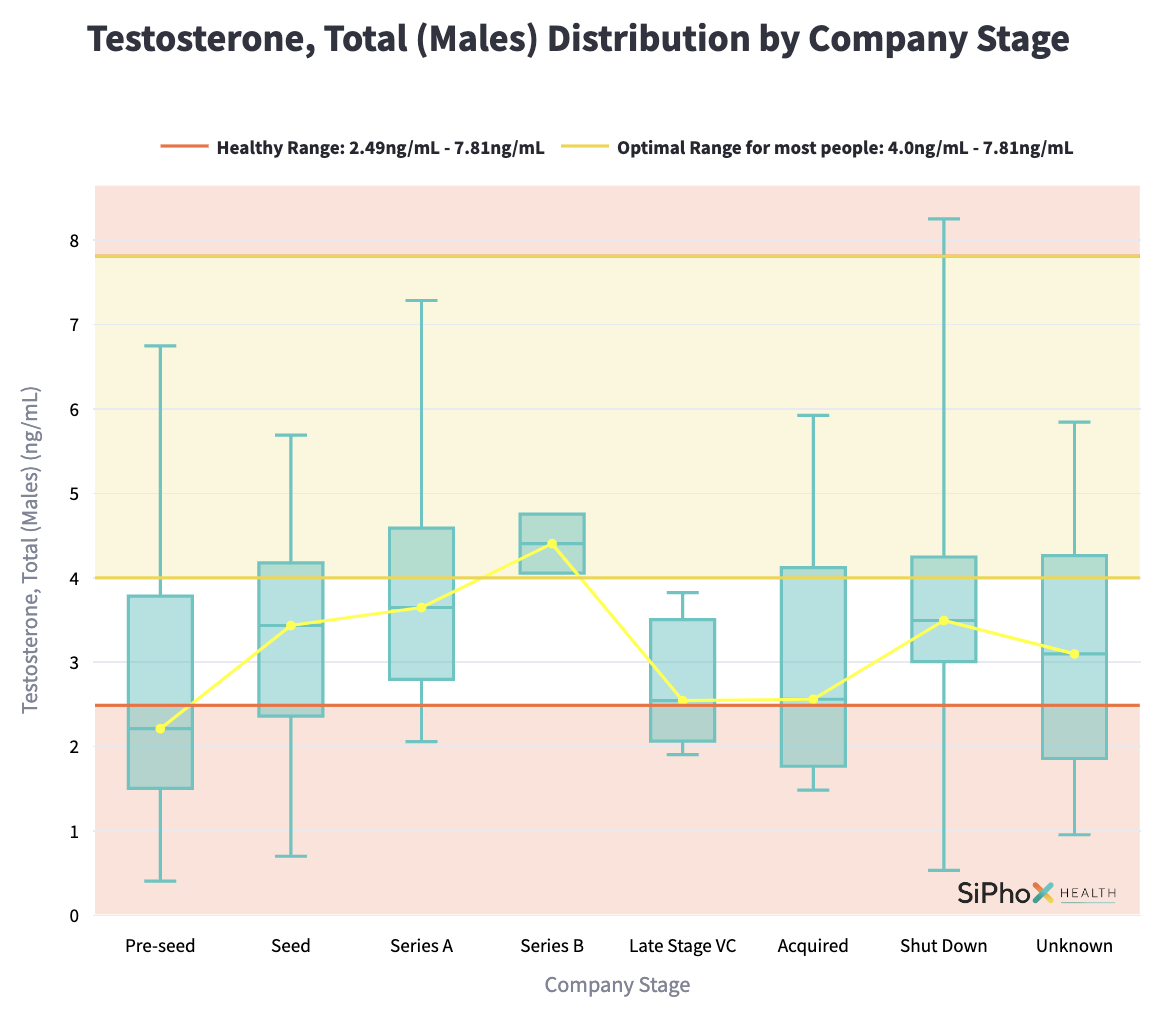}
    \captionsetup{width=0.7\linewidth} 
    \caption{Age-normalized total testosterone distribution for males across company stages. Error bars represent interquartile ranges.}
    \label{fig:testosterone}
\end{figure}

Comparing these values to the healthy range (2.49--7.81~ng/mL) and the optimal range (4.00--7.81~ng/mL), which was developed by SiPhox Health, for testosterone, we observe that founders at the pre-seed and acquired stages had median levels below the healthy range. Seed and Series A founders were within the healthy range but below optimal levels. Only at the Series B stage did the median testosterone level enter the optimal range, although statistical significance could not be established due to the small sample size (figure~\ref{fig:testosterone}).

The Spearman correlation analysis for testosterone levels, excluding the Shutdown and Unknown stages, showed a positive and significant trend across company stages ($p$\,=\,0.035).

\subsection{Cortisol Levels by Company Stage}

Cortisol, a hormone associated with stress and known to inversely affect testosterone levels, remained relatively stable through the Series B stage and then spiked at the late-stage VC phase (\autoref{table2}). The median cortisol level increased from 6.03~nmol/L at Series B to 8.44~nmol/L at the late-stage VC phase, representing a 39.9\% increase. Due to small sample sizes in the Series B ($n$\,=\,2) and late-stage VC ($n$\,=\,3) stages, statistical significance could not be assessed for these changes.

\begin{table}[t] 
\centering 
\caption{Median Cortisol Levels Across Company Stages} 
\label{table2} 
    \begin{tabular}{lccc} 
    \hline 
    \textbf{Company Stage} & \textbf{Median Cortisol (nmol/L)} & \textbf{\% Change from Previous Stage} & \textbf{Sample Size ($n$)} \\
    \hline 
    Pre-seed       & 7.36 & --- & 34 \\
    Seed           & 5.94 & --19.3\% & 29 \\
    Series A       & 7.46 & +25.6\% & 11 \\
    Series B       & 6.03 & --19.2\% & 2 \\
    Late Stage VC  & 8.44 & +39.9\% & 3 \\ 
    Acquired       & 7.06 & --16.4\% & 6 \\ 
    Shut Down      & 6.51 & --- & 10 \\ 
    Unknown        & 6.75 & --- & 12 \\ 
    \hline 
    \end{tabular} 
\end{table}

\begin{figure}[H]
    \centering
    \includegraphics[width=0.75\linewidth]{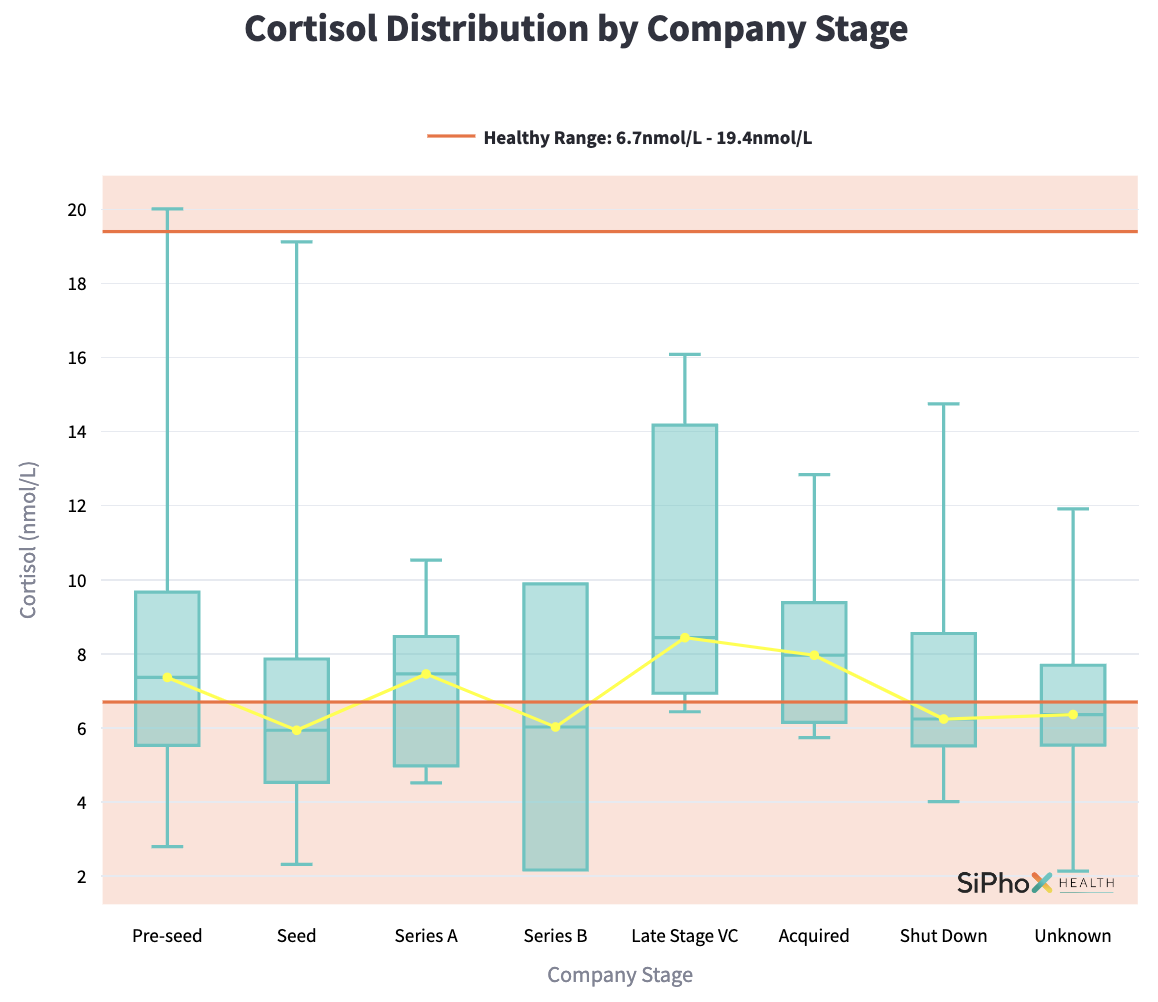}
    \captionsetup{width=0.7\linewidth} 
    \caption{Age-normalized cortisol distribution for males across company stages. Error bars represent interquartile ranges.}
    \label{fig:cortisol}
\end{figure}

The optimal range for cortisol is 6.7--19.4~nmol/L. Founders at the seed and Series B stages had median cortisol levels below the optimal range, whereas those at the pre-seed, Series A, and late-stage VC phases were within the optimal range (Figure~\ref{fig:cortisol}).

\subsection{Dehydroepiandrosterone Sulfate (DHEA-S) Levels by Company Stage}

\begin{table}[t] 
    \centering 
    \caption{Median DHEA-S Levels Across Company Stages} 
    \label{tab} 
    \begin{tabular}{lccc} 
        \hline \textbf{Company Stage} & \textbf{Median DHEA-S (µg/dL)} & \textbf{\% Change from Previous Stage} & \textbf{Sample Size ($n$)} \\ 
        \hline 
        Pre-seed       & 225.99 & --- & 34 \\ 
        Seed           & 183.82 & --18.7\% & 29 \\ 
        Series A       & 220.95 & +20.2\% & 11 \\
        Series B       & 220.50 & --0.2\% & 2 \\
        Late Stage VC  & 247.27 & +12.1\% & 3 \\ 
        Acquired       & 251.04 & +1.5\% & 6 \\ 
        Shut Down      & 312.29 & --- & 10 \\ 
        Unknown        & 168.58 & --- & 12 \\ 
        \hline 
    \end{tabular} 
\end{table}

DHEA-S, a precursor hormone to testosterone, exhibited a slight decrease between the pre-seed and Series B stages, followed by an increase at the late-stage VC phase (\autoref{tab}). The median DHEA-S level decreased by 2.4\% from 225.99~µg/dL at the pre-seed stage to 220.50~µg/dL at the Series B stage. At the late-stage VC phase, the median DHEA-S level increased to 247.27~µg/dL, a 12.1\% rise from the Series B stage.

\begin{figure}[H]
    \centering
    \includegraphics[width=0.75\linewidth]{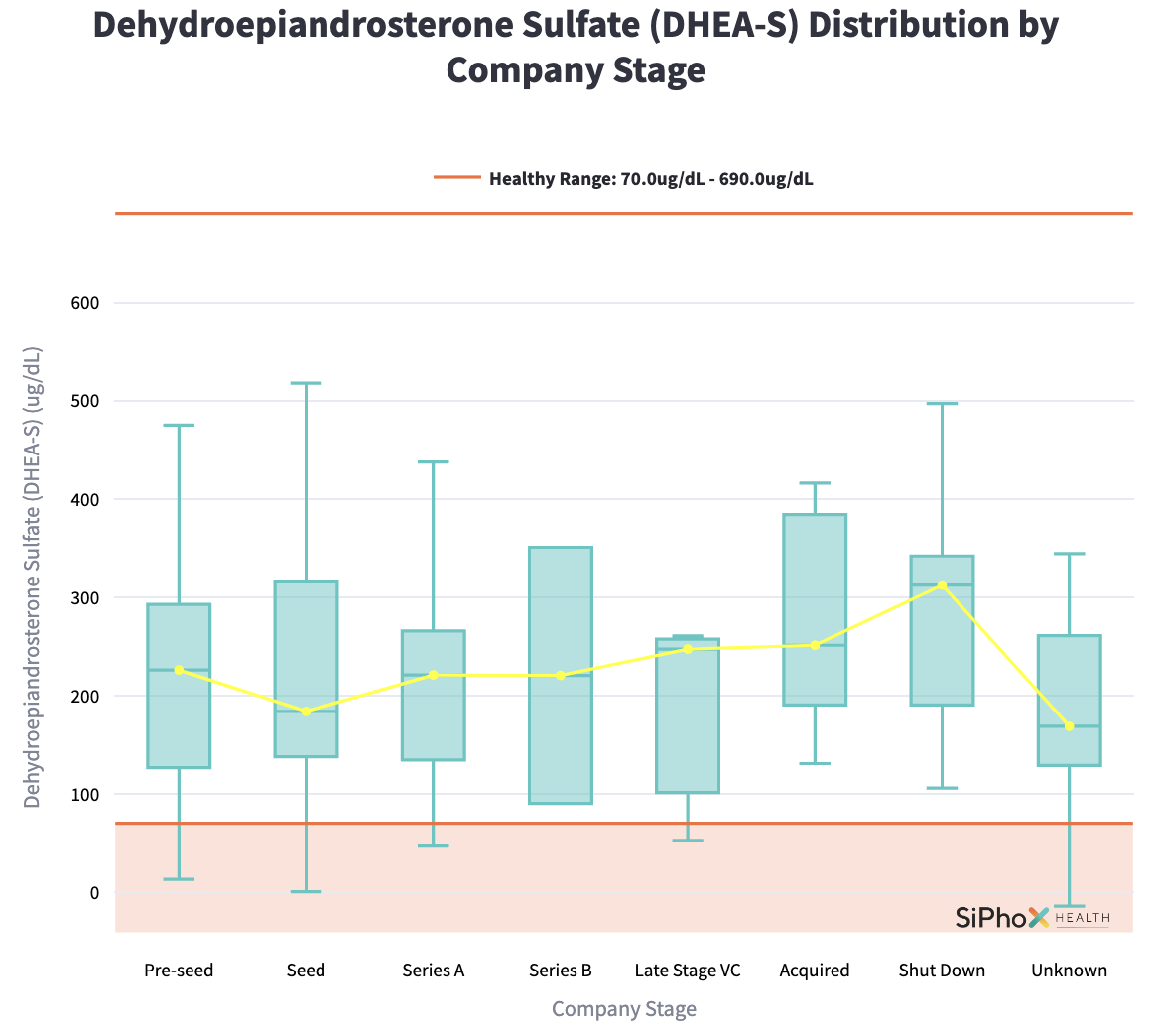}
    \captionsetup{width=0.7\linewidth} 
    \caption{Age-normalized DHEA-S distribution for males across company stages. Error bars represent interquartile ranges.}
    \label{fig:dheas}
\end{figure}

The optimal range for DHEA-S is 70--690~µg/dL. All median values across the company stages fall within this range (Figure~\ref{fig:dheas}). 

\section{Discussion}\label{sec12}

Understanding the relationship between startup lifecycle stages and founders' hormonal health is essential, as it may provide insights into the physiological underpinnings of entrepreneurial performance and well-being. Our data raises intriguing questions such as: Do founders at earlier stages exhibit lower testosterone levels due to heightened stress, long work hours, and reduced sleep? Are higher testosterone levels associated with successful fundraising and company progression, possibly reflecting increased confidence?

The hormonal trends observed in this study indicate a complex interplay between testosterone, cortisol, and DHEA-S levels as founders navigate different stages of company growth. The significant increases in testosterone levels between the pre-seed, seed, and Series A stages may reflect heightened feelings of success, confidence, and reduced stress as companies grow and secure more funding. This pattern suggests that early funding stages are associated with positive psychological and physiological changes. However, the subsequent decline in testosterone and the concomitant rise in cortisol at the late-stage VC phase could reflect heightened pressures related to scaling operations, meeting investor expectations, or preparing for potential exits.

Although this interpretation appears likely, it is also possible that founders with inherently higher testosterone are more likely to reach later stages of funding. This would only be possible to prove with a longitudinal study, however, undertaking such a study may make sense for Venture Capitalists or others who might be interested in finding new and powerful predictors of startup success. 

The cortisol data add another layer to this complex picture. The spike in cortisol levels observed at the late-stage VC phase aligns temporally with the decline in testosterone, supporting the well-documented antagonistic relationship between these two hormones~\cite{mehta2010testosterone}. High cortisol levels can inhibit testosterone production, suggesting that stress in these later stages could contribute directly to the reduction in testosterone seen in founders at this phase.

As Mehta and Josephs (2010) articulate in their seminal paper, "In the domains of leadership (Study 1, mixed-sex sample) and competition (Study 2, male-only sample), testosterone was positively related to dominance, but only in individuals with low cortisol"~\cite{mehta2010testosterone}. This dual-hormone hypothesis underscores the impact of stress on hormonal balance, indicating that the heightened pressures associated with scaling operations, meeting investor expectations, and navigating complex environments during late-stage rounds may suppress testosterone levels. Consequently, this hormonal suppression could influence founders' physiological well-being, decision-making abilities, and overall performance and dominance during critical phases of their startup's lifecycle. Conversely, Mehta and Josephs' findings support the notion that the combination of low cortisol and progressively increasing testosterone observed in founders from pre-seed through Series B stages may causally contribute to heightened dominance behaviors.

Understanding this interplay between cortisol and testosterone provides valuable insights into the physiological challenges faced by entrepreneurs, highlighting the importance of stress management and hormonal health in fostering sustained entrepreneurial success and dominance.

In contrast, the decline in DHEA-S from the pre-seed to Series B stages diverges from the overall trend observed for testosterone, suggesting that factors beyond DHEA-S availability are likely influencing testosterone production during these stages.

It is worth noting, that founders with "Shut Down" companies had a huge variance in age-adjusted Testosterone levels, which may be an artifact of the small dataset, but also may suggest that shutting down their startup can be either demoralizing or a huge weight off their shoulders for founders.

Further research is needed to explore the causal relationships and underlying mechanisms driving these hormonal changes. Factors such as sleep patterns, physical activity, and stress management strategies should be considered to fully understand their impact on hormonal profiles during different stages of company development. We also recognize that our study is limited by the small sample sizes in some of the later stages, and we intend to repeat this study with a larger sample size to increase statistical power and reliability.

\subsection{Future Studies}\label{future}

The data suggests several follow-on studies. Analyzing biomarker results against actual funding amounts, rather than categorizing by company stage alone may yield cleaner correlations since funding stages are ill-defined. 
Additionally, participants could be tracked longitudinally to understand whether testosterone levels increase as founders progress through rounds. This longitudinal analysis would help clarify the causal direction of the observed relationship—whether higher testosterone is a factor that helps founders advance to Series B or if successful progress leads to increased testosterone. Potential interventions, such as enclomiphene treatment or lifestyle modifications aimed at boosting testosterone, should be explored to determine if mechanistically increasing testosterone levels has a meaningful impact on the likelihood of advancing through funding rounds.
The current findings provide a starting point for understanding how physiological changes might reflect or even influence startup dynamics, and we are excited to build on this foundation with expanded studies in the future.

\backmatter

\section*{Declarations}

SiPhox Health is a for-profit at-home diagnostics Y Combinator-funded startup.\newline \newline The authors excluded SiPhox Health founder data from this dataset, but it suffices to say that the company completed its Series A funding round in 2023. 


\bibliography{sn-bibliography}% common bib file

\end{document}